# Mitigating the Effects of Ransomware Attacks on Healthcare Systems

Sreejith Gopinath
*Computer Science and Engineering*
*NYU Tandon School of Engineering*
New York, USA
sg7192@nyu.edu

Aspen Olmstead
*Computer Science and Engineering*
*NYU Tandon School of Engineering*
New York, USA
aspeno@nyu.edu

*Abstract - Healthcare information systems deal with a large amount of Personally Identifiable Information related to patients like dates of birth and social security numbers, patients' health information and history, and financial information like credit card details and bank accounts. Most healthcare institutions purchase information systems from commercial vendors and have minimal in-house expertise required to maintain these systems. Most institutions lack the expertise required to research evolving threats and maintain a tough security posture. We propose a risk transference-based system architecture that moves sensitive data outside the system boundary, into data stores that are managed with stringent and efficient security protocols.*

## I. Introduction

The Cybersecurity firm *Cybersecurity Ventures* projects in their report that the costs associated with business disruption and data losses will rise to an estimated $265 billion over the next decade [1]. The *US Department of the Treasury* notes in their 2021 report that ransomware-related transactions of around $590 million occurred in the first half of 2021 alone, a 40% increase over the entire prior year's transactions [2]. Although cryptocurrency transactions do not ensure perfect anonymity, the enhanced anonymity afforded by cryptocurrency transactions has also contributed to the audacity of these attacks, with an estimated $5.2 billion worth of cryptocurrency transactions related to ransomware attacks in the first half of 2021 alone.

Compared to ransomware attacks against other industries, attacks against healthcare systems have garnered heightened visibility due to the humanitarian costs. Attacks against healthcare organizations have steadily escalated over recent years, especially since the beginning of the COVID-19 pandemic. Apart from causing disruptions that can delay research outcomes and life-saving treatments, malware attacks also threaten to expose Personally Identifiable Information (PII) of patients and providers externally. In addition, health conditions and other medical records tied to the PII could be disclosed. A wide variety of healthcare software systems are in use, and they all differ in the way they store and process sensitive data, preventing the adoption of any one single standard or protocol in order to combat the problem. The intense strain on healthcare systems caused by the effects of the pandemic has exacerbated the problem.

The main thrust of this work is the exploration of alternate mechanisms of storing PII and mappings between PIIs and other data to make systems resilient to ransomware attacks. In Section II, we discuss some recent related research and the pain points that we attempt to mitigate. In Section III, we discuss some approaches that have been proposed in the past to try and address this problem. In Section IV, we perform threat modeling for a common implementation of a healthcare system and provide some commentary on some ways that potential bad actors may compromise the system. In section V, we hypothesize an approach that we believe will render the healthcare application and storage system unattractive to would-be attackers by keeping sensitive data and mappings outside the system boundary, in specialized data storage systems that adhere to stringent security protocols. In section VI, we outline our plan to build a prototype of our proposed solution to aid in evaluation and present the metrics that we will evaluate to understand the efficacy of our proposed solution. In section VII, we discuss the prototype system implementation in some detail. In section VIII, we describe the comparative evaluation of our proposed solution and discuss the cost and resilience trade-offs. In section IX, we conclude our work on this topic and present some thoughts as to the future of this work. As far as we know, no other work has proposed mitigating the effects of a successful ransomware attack by decoupling the data from the logic that processes the data and storing data without accompanying context.

## II. Related Research

Zaki et al., in their work on detecting and preventing a ransomware attack [3], define a Next-Generation Firewall as one that can analyze and evaluate network packets in real-time, detect an attack as it is underway, and prevent an attack by blocking this traffic. This solution works when the intrusion detection system can see malicious packets in real-time; however, without significant continuous learning capabilities, bad actors may craft packets that can evade the intrusion detection systems and infiltrate the network.

Thamer et al., in their study of known attacks [4], discuss how commonly known attacks affect a typical healthcare information system. The known attacks include email phishing campaigns, exploit kits, and watering hole attacks. They discuss several remedies to combat ransomware by utilizing blockchain technology to make patient data immutable, using machine learning technologies to detect attacks against the network, and using a software-defined network to attempt to block malicious traffic. The main shortcoming of this approach is that the data is still described as within the system boundary and the security of the data depends on the effectiveness of the algorithms that detect an attack that is underway.

Okereafor et al., in their work on surveying recent ransomware attacks and their effects [5], discuss the cybersecurity risks in the healthcare industry in the context of the COVID-19 pandemic, the effects of cybersecurity attacks on healthcare information systems, and suggest a few different approaches to prevent attacks that could compromise sensitive healthcare data like institutionalizing a cybersecurity-minded culture and implementing multi-factor authentication. However, the work stops short of proposing concrete system architectures or specific methods to prevent or remedy a cybersecurity attack.

Elsayed et al., in their work exploring the effects of attacks on patient data privacy [6], describe a few cryptography-based solutions to protect sensitive patient data from breaches or leaks, but do not address the ransomware problem and how sensitive data could be held to ransom and rendered inaccessible for legitimate business purposes. Cryptography is a strong defense against unauthorized access to data, and we use cryptography in our proposal to eliminate the threats posed by various well-known attacks.

Ahmed et al., describe several security challenges prevalent against this sector and propose a set of cybersecurity metrics that could be used to improve the security posture of the organization [7]. We have adapted some of the metrics proposed into metrics that we use to evaluate the effectiveness of our own approach.

Abu Ali et al., in their work, discuss why security should be enhanced in healthcare systems and explore some ways of doing so [8]. The importance of securing healthcare systems as discussed by Abu Ali et al. have only been made more urgent in the context of the COVID-19 pandemic.

Javid et al., discuss the security aspect of transmitting sensitive patient data in their work focused on facilitating analytics on patient information [9]. Their work focuses more on securing data in flight. Our proposed approach agrees with the need to secure data in flight but goes a step further in discussing the need to secure data at rest too.

The approach described in this paper goes a step further by proposing a data storage and retrieval architecture that keeps sensitive data outside the system boundary. Encrypted references to the data are maintained within the system boundary so that malicious actors will not access sensitive information just by infiltrating the network. This renders the healthcare information system unappealing as a target to malicious actors (since malicious actors, even if they can infiltrate the network, will not have access to the data).

### III. OTHER SOLUTIONS PROPOSED IN THE PAST

The problem of ransomware attacks has been around for a long time and several solutions have been considered in the industry at large. Several of these solutions have been considered for application in the healthcare industry too. As part of previous efforts, healthcare systems have moved towards implementing better security protocols and overall maintenance of a tough security posture. However, cyber threats are constantly evolving and staying abreast of them requires considerable investment and energy. User trainings and sensitization campaigns have served to thwart low-sophistication attacks, but as in the previous case, do little to protect against a highly evolved attack. Adopting IaaS and PaaS offerings in their software implementations have worked to eliminate several threats but the downside is that these are getting increasingly complex and require significant in-house IT expertise to implement and maintain.

TABLE 1: PREVIOUS SOLUTIONS ATTEMPTED

| Solution | Pros | Cons |
|---|---|---|
| *Implement better security protocols and best practices.* | As expected for a general software system. | Additional effort for employees whose primary function is not to deal with application security. <br><br> Security postures must be evaluated continuously and adjusted; incurs more overhead. |
| *Implement continuous user training and sensitization campaigns.* | Helps users adopt a security mindset. | Threats evolve at a rapid pace and training will most likely not be able to keep up. <br><br> It must be accepted that users may slip up. One is all it takes. |
| *Increasing use of IaaS and PaaS solutions.* | Works to eliminate some of the threats identified in the threat model. | There are multiple points at which a bad actor could gain access into the system boundary and get access to storage or storage APIs. |

### IV. THREAT MODELING

The threat modeling section discusses the threat modeling of a standard architecture of healthcare systems in common use. The specific systems we are discussing here allow front-office staff to store PII like name, date of birth, address, social security numbers, etc. from patients and medical professionals (doctors, nurses, physicians' assistants) and store them in the system. Following a medical visit, the medical professional stores medical data including vital measurements, diagnosed medical conditions and prescribed medications in the system. Analytical components join the PII with medical data to develop insights. These tools include statistical analysis and machine learning modules. Patients utilize personal computing and mobile devices to access their data. Fig 1 shows a threat model for a standard healthcare system implementation. The model divides the system into various *zones*, each of which comes with a different threat potential.

The *Cloud Gateway Zone* comprises the storage layer and the APIs that allow access to the data. This zone is a generalized boundary covering healthcare systems that persist their data on-prem or in cloud offerings like GCP or AWS. This zone encompasses the data storage solutions and usually have some sort of APIs to regulate the ingress and egress of data into the zone. In this diagram of a threat model, storage APIs regulate access to the stored data.

*The Local User Zone* comprises the hospital's device zone (devices/terminals used by hospital employees) that are used to persist sensitive information, the authentication modules and the modules that perform analytical operations over the persisted data.

*The Device Zone* includes remote devices that patients, partners, and regulators use to access data in the hospitals' systems. This zone is outside the system boundary and is not integral to the proper functioning of the hospital healthcare system.

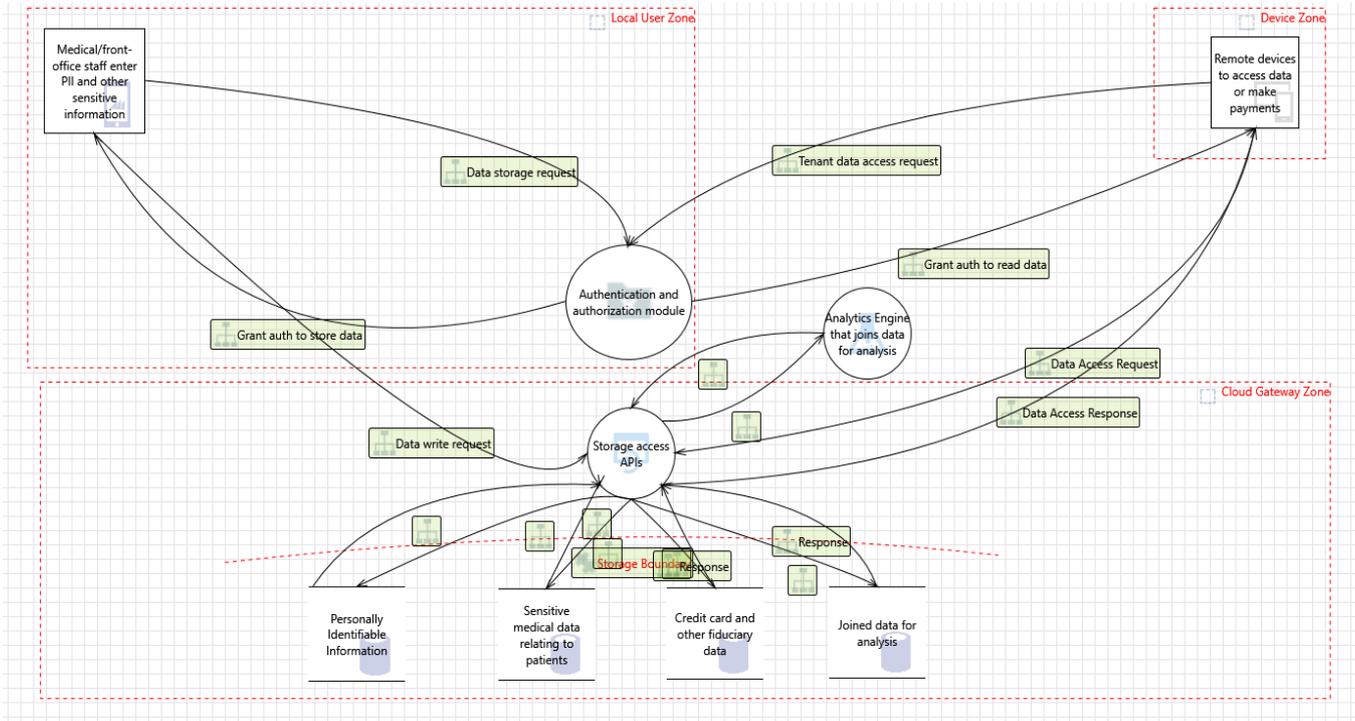

Fig 1: Threat model for a standard healthcare system implementation

Bad actors could attack the system at various points. We enumerate a few possibilities here:

A. A bad actor could impersonate an employee of the hospital (steal password or private key, etc.) and write/read sensitive data.
B. A bad actor could compromise the authentication module and gain access to the storage layer.
C. A bad actor could exploit a vulnerability in the system that allows a patient's devices to be used to perform unauthorized data operations.

The data stored in the system could be compromised in any of the scenarios presented here. Our work focuses specifically on PII and sensitive medical data and the attacks that expose or hold the data to ransom.

## V. HYPOTHESIS

The threat model shown above makes it evident that there is more than one layer at which a potential attacker could launch their attacks to (i) Steal PII stored in a healthcare system (ii) Hold the data in storage to ransom. Several streams of research are currently exploring how to deny entry to attackers into software systems that hold sensitive data. Our hypothesis here attempts to address the problem from a different angle. Rather than concentrate on ways to keep attackers out of the software system, we assume that it is impossible to secure an application completely and that attackers will eventually find some way to enter the system. Therefore, the solution must render the target unattractive to attackers because they couldn't extract any sensitive data from the system. The ransomed system could simply be abandoned, and a new instance instantiated in its place with minimal disruption to the business.

The proposal put forth by us in this work is that sensitive data relating to patients are not stored within the healthcare information system but instead within a central repository outside the system boundary. This repository is a third-party service not managed by the healthcare system at all. The repository could provide similar services to several other healthcare systems or a varied portfolio of clients. The repository will have its own security and administrative policies, using those strategies to hold data at the highest security levels with several layers of redundancy. By shifting the burden of security from organizations that lack the expertise to thwart sophisticated attacks to organizations whose bread-and-butter business is the security and availability of confidential information, this work explores the possibility of making organizations unattractive to bad actors as possible.

## VI. PROPOSED EVALUATION AND KEY METRICS

The proposal in this paper proposed building a proof-of-concept system that utilizes a remote repository (outside the system boundary) to store simulated sensitive data. The remote repository is assumed to employ advanced protocols to ensure safety and authorized access to the data stored in the repository. These kinds of repositories are widely used today as password managers. We, therefore, assume password managers to be out of scope for this work. Instead, our work will focus on designing a healthcare system that utilizes such a repository to store sensitive data and discuss the mechanisms to authenticate and ensure authorized access to the data. We will also discuss how systems under ransom may be recovered quickly without needing to engage with bad actors.

We will propose ways to evaluate the success of our proposed solution by building a prototype of the system that first stores the data as unencrypted files on disk, then as encrypted files on disk, and then in a database, all within the same healthcare system boundary.

The prototype will subsequently be modified to store the data in a separate storage system outside the healthcare system boundary. This storage system will be simulated, and we will assume that it has the same security protocols as several systems in use for other purposes (password managers, etc.) today.

We will discuss the performance implications of each choice and the ease of operations recovery following a successful breach of the healthcare system.

TABLE 2: PROPOSED EVALUATION CRITERIA BY FUNCTIONALITY

| Data Storage Approach | Vulnerability | (D)Encryption Costs on access | Data transfer costs | (D)Encryption costs at storage | Usage complexity score | Attack sophistication involved |
|---|---|---|---|---|---|---|
| *Unencrypted local filesystem.* | Data can be leaked/ransomed. | None | No | None | Low | Low |
| *Encrypted local filesystem.* | Data can be ransomed. | Low | No | Low | Medium | Low |
| *Unencrypted remote filesystem.* | Data can be leaked /ransomed. | Low | Yes | None | Medium | Medium |
| *Encrypted remote filesystem.* | Data can be ransomed. | Low | Yes | Low | Medium | Medium |
| *Local database.* | Data can be leaked/ransomed. | Low | No | Medium | Low | Medium |
| *Remote database.* | Data can be leaked. | Low | Yes | Medium | Medium | High |
| *Specialized, remote data storage (our proposal).* | **Data cannot be leaked or ransomed.** | **Medium** | **Yes** | **High** | **High** | **Very High** |

The table above shows that the higher the attack sophistication required to compromise the security and gain access to the data or place data under ransom, the better. The higher the usage complexity score, the higher the cost paid for security. We should be striving for a solution that requires as high an attack sophistication score as possible while keeping in mind the costs incurred. Table 2 shows the different computational and network transfer costs applicable to each approach.

The table above shows that increased computational and network transfer costs are the price we expect to pay for having a secure data storage solution that renders a healthcare system unattractive to would-be bad actors.

VII. PROTOTYPE SYSTEM IMPLEMENTATION

We describe the implementation of a prototype of the proposed system with seven different data storage approaches:

A. Data stored in encrypted/unencrypted local files.
B. Data stored in encrypted/unencrypted remote files.
C. Data stored in a local/remote database.
D. Our proposed approach of storing data in simulated remote storage for sample PII data.

We ran experiments to compare performance over each class of solutions and ponder the attack sophistication required to compromise each system and gain access to sensitive data.

A. Hardware

The prototype healthcare system was deployed on a Lenovo laptop (the application host) with AMD Ryzen™ 5 5500U Processor (6 Cores / 12 Threads, 2.10 GHz), 16GB RAM and 802.11 wireless networking, running Ubuntu 20.04. The remote data storage approaches utilizing encrypted and unencrypted remote file storage, remote database, and the proposed simulated data store were all deployed on a Raspberry Pi 4 computer. The Raspberry Pi (the data host) had a Broadcom BCM2711, Quad-core Cortex-A72 (ARM v8) 64-bit SoC, 1.5GHz Processor, and 8GB RAM, running the Raspbian Buster OS.

B. Software

Encrypted and unencrypted files were all on the Linux file system (whether local or remotely on the Pi). For the database approach, we used SQLite.

The prototype healthcare system is implemented as a Spring boot application written in Java and has APIs supporting the creation, updating, deletion, and listing of patient records. Each live instance of the healthcare application gets its own identifier.

The proposed data store is simulated as a data store with a simple key-value store at its core and has APIs to store and list data sent to it.

C. Implementation of the healthcare application

The application has the following APIs to support operations:

1. Create

The Create operation is used to create new patient records (PII, medical data and financial information). On a successful write, the proposed data storage system returns a unique key for the written data.

2. Read

The Read operation is used when patient details need to be retrieved (by a healthcare professional, patient portal, etc.). Data is deemed to be lost if the unique reference to the data is lost.

3. Update

The Update operation is used when patient details need to be updated (for example, if the patient has a new address).

4. Delete

The Delete operation is used when patient details need to be deleted from the system.

To experiment with the file system approaches, we created one directory to hold personal data and another to hold credit card information. The healthcare application stored patient data in files named with Patient IDs in their respective folders.

In order to experiment with the database approaches, we set up a SQLite database locally on the application host and the Raspberry Pi 4; JDBC connections were used to interact with the database.

The paper describes how, for the approach proposed, the healthcare application system registers itself to a central registry on initialization. It receives a unique identifier (instance ID) and authorization code (license key) on successful registration. These values are "salted" by appending the license key to the instance ID and the resultant value encrypted using a high number of iterations of AES-256 encryption. This process yields the encryption key used to encrypt sensitive data before it is sent to the data store.

The application host and data hosts communicate over HTTPS to prevent man-in-the-middle attacks and eavesdropping.

D. Implementation of the proposed data store

The implementation of the proposed data store is similar to that of the healthcare application; the data store is also implemented as a Spring boot application in Java. At the heart of this system is a massive key-value store that holds all the data. The data store application exposes just two APIs as described below:

1. Create

The Create operation is used to persist new records into the data store. When the data store application receives a piece of data, a UUID is generated and used as a key to store the encrypted data. The UUID is returned to the healthcare system to use as a lookup value. The UUID is then stored in the list of references that will be used to recover the system in case it is placed under ransom and needs to be abandoned.

2. Delete

The Delete operation is used to delete records from the healthcare system. If a record is found matching the provided key, that record is purged from the system.

## VIII. EVALUATION OF ALL APPROACHES

We describe how we implemented a prototype of a healthcare system and data stores for several approaches (filesystem, database, dedicated data store), and we deployed them to their respective hardware platforms. In addition, both hardware platforms were networked over a 802.11ac wireless link.

Ten million rows of synthetic data were fed into the system. This data translates into 10 million rows of patient PII. The data was delimited by question marks and included the following three fields: date_of_birth, social_security_number and address. Ten million rows of financial information were persisted too; the format of each row was similar with a question mark delimiter and the following three fields: credit_card_number, expiration_date, auth_code.

We simulated a ransomware attack by the following means:

A. Removing the application host from the wireless link, OR
B. Removing the data host from the wireless connection.

We did not simulate attacks on both systems simultaneously, as the chances of it happening in the real world are slim.

Evaluation of resilience to attack, costs of use and ease of recovering a ransomed system

We discuss how, for each approach, we attempted to measure the resilience to attack, computational and network transfer costs, and ease of recovering a ransomed system. We average out the costs over ten million insert and 1 million fetch operations. Those findings are presented in Table 4.

The attacks were simulated by taking the specified hosts off the wireless link. Since we had a copy of the executables for both applications (healthcare system and data store), we were able to spin up an instance of the system that was being tested to check if there was any way to resume the state of the system immediately before the attack.

TABLE 3: EVALUATION OF PROPOSED SOLUTION AGAINST OTHER APPROACHES OVER COST CRITERIA

| Approach | Encryption time (msec) - Create | Decryption time (msec) - Fetch | Network time (usec) | Resilience to attack | Ease of data recoverability |
|---|---|---|---|---|---|
| Unencrypted local filesystem. | 0 | 0 | 0 | Low | Potential data loss; system could not be recovered. |
| Unencrypted remote filesystem. | 0 | 0 | 118 | Low | Potential data loss; system could not be recovered. |
| Encrypted local filesystem. | 1.22 | 1.38 | 0 | Moderate | Data was presumed safe; system could not be recovered. |
| Encrypted remote filesystem. | 1.22 | 1.38 | 118 | Moderate | Data was presumed safe; system could not be recovered. |
| Local database. | 0 (database times transparent to users). | 0 (database times transparent to users). | 0 | Low | Data was presumed safe; system could not be recovered. |
| Remote database. | 0 (database times transparent to users). | 0 (database times transparent to users). | 118 | High | Data was presumed safe; system could be recovered from database snapshots. |
| **Storage approach proposed by this work.** | 1.22 | 1.38 | 118 | Very High | A new instance of the system was spun up with the same identifier, the "salted" key was recomputed and the list of data references was read from the data store. The above two operations were completed in less than 2 minutes and the healthcare system was back in business! |

## IX. Conclusion

The work that we have done shows that the architecture proposed in this paper can deter attacks to a great extent, as evidenced by the results of the experiments above. In addition, we have shown how storing sensitive information without accompanying context renders the data less appealing to bad actors. Finally, we have also demonstrated how easy it is for authorized users to recover a system that has been placed under ransom. This demonstration should motivate victims to not negotiate with bad actors seeking a ransom in return for freeing the system and instead attempt to recover the system with minimal business disruption.